\documentclass[12pt,epsf]{article}

 \usepackage{epsfig}
 \usepackage{times}
 
 \usepackage{amsmath}
 \usepackage{amssymb}

\newcommand{\be}{\begin{equation}}
\newcommand{\ee}{\end{equation}}
\def\({\left (}
\def\){\right )}
\def\[{\left [}
\def\[{\right ]}

\begin{document}
\begin{titlepage}
\bigskip
\rightline{}
\rightline{hep-th/0602003}
\bigskip\bigskip\bigskip\bigskip
\centerline {\Large \bf {Gravity Dual of Gauge Theory on $S^2\times S^1\times R$}}
\bigskip\bigskip
\bigskip\bigskip

\centerline{\large Keith Copsey and Gary T. Horowitz}
\bigskip\bigskip
\centerline{\em Department of Physics, UCSB, Santa Barbara, CA 93106}
\centerline{\em keith@physics.ucsb.edu, gary@physics.ucsb.edu}
\bigskip\bigskip

\begin{abstract}
We (numerically) construct new static, asymptotically AdS solutions
where the conformal infinity is the product of time and $S^2\times S^1$. There always exist a family of solutions in which the $S^1$ is not contractible and, for small $S^1$, there are two additional  families of solutions in which the $S^1$ smoothly pinches off. This shows that (when fermions are antiperiodic around the $S^1$) there is a quantum phase transition in the  gauge theory as one decreases the radius of the $S^1$ relative to the $S^2$.   We also compare the masses of our solutions and argue that the one with lowest mass should minimize the energy among all solutions with conformal boundary $S^2\times S^1\times R$. This provides a new positive energy conjecture for asymptotically locally AdS metrics. A simple analytic continuation produces AdS black holes with topology $S^2\times S^1$.
\end{abstract}
\end{titlepage}

\baselineskip=16pt
\setcounter{equation}{0}

 \section{Introduction}
 
There has been extensive work exploring the consequences of the AdS/CFT correspondence. The most familiar case is the duality between four dimensional ${\cal N} =4$ super-Yang-Mills and Type IIB string theory on spacetimes which asymptotically approach  $AdS_5\times S^5$. It is sometimes assumed that the dual description of the vacuum state is given by just taking the bulk metric to be $AdS_5\times S^5$ everywhere. However, this is only true if the gauge theory is on $S^3$ or $R^3$. If instead,  we consider the gauge theory on $S^1\times R^2$ or $T^3$ and require fermions to be antiperiodic around a circle, there is a lower energy solution called the AdS soliton \cite{Horowitz:1998ha,Witten:1998zw}.  The fact that this solution has less energy than periodically identified $AdS_5$ can be related to the existence of  negative Casimir energy in the dual field theory.

In this paper we consider the dual description of the vacuum state of the gauge theory on
$S^2\times S^1$. Since  the $SO(6)$ symmetry of the scalars is unaffected, the bulk geometry should be the product of $S^5$ and a five dimensional solution to Einstein's equation with negative cosmological constant.  There is no way to periodically identify $AdS_5$ to obtain a conformal boundary which is $S^2\times S^1 \times R$,  so we must construct a nontrivial solution.  We numerically construct a family of static solutions with the right asymptotic behavior.  Properties of these solutions depend on the ratio of the radius of the $S^1$ to the radius of the $S^2$. When this ratio is large there is a unique static solution with topology $R^4\times S^1$. When it is small, there are three different solutions with the same conformal infinity. In addition to the solution with topology $R^4\times S^1$, there are two solutions in which the circle smoothly pinches off, producing a space with topology $R^3\times S^2$. Since the $S^1$ is contractible, there is a unique spin structure for fermions and they must be antiperiodic. This shows that there is a quantum phase transition in the strongly coupled gauge theory as one shrinks the size of the circle. 

Our results are similar to what has been found for gauge theories on $S^n$ at finite temperature \cite{Witten:1998zw}.\footnote{There is an extensive literature on analogous phase transitions for gauge theories on tori. See, e.g., \cite{Surya:2001vj,Aharony:2005ew}  and references therein.}  This is described by a Euclidean functional integral  on $S^n\times S^1$ where the $S^1$ is now Euclidean time. At low temperatures, corresponding to large $S^1$, there is a single bulk solution representing a gas of particles in AdS. At high temperatures, there are two additional solutions corresponding to the analytic continuation of large and small black holes. However, even though our results are similar,  there is a crucial difference. We are considering the zero temperature ground state of  Lorentzian gauge theories on $S^2\times S^1$. Our bulk geometries are globally static and Lorentzian. If we wanted to investigate these theories at finite temperature, we would have to consider Euclidean solutions with asymptotic geometry $S^2\times S^1 \times S^1$.

In addition to providing insight into strongly coupled gauge theories, there is another motivation for this work.  In asymptotically flat spacetimes with standard Kaluza-Klein boundary conditions,  there are vacuum solutions with arbitrarily negative energy \cite{Brill:1989di,Brill:1991qe}. These solutions are sometimes called Kaluza-Klein ``bubbles of nothing" \cite{Witten:1981gj} since the circle pinches off at a nonzero radius, producing a minimal area sphere. There is no spacetime inside this radius. The analog of Kaluza-Klein boundary conditions for AdS are spacetimes with a conformal infinity $S^2\times S^1\times R$. If there were solutions with arbitrarily negative energy in this case,  it would pose a serious problem for the AdS/CFT correspondence since the gauge theory on $S^2\times S^1$ should have a stable ground state.  We provide evidence that the energy is always bounded from below in AdS.  Our solutions in which the circle pinches off are like static bubbles.\footnote{Although Witten's bubble expands rapidly, static bubbles also exist in asymptotically flat spacetime, e.g., the product of time and the Euclidean Schwarzschild metric.} For small $S^1$, one of the static bubbles has the lowest energy among the three static solutions.  We conjecture that this solution has the lowest energy among  all solutions with this asymptotic structure. This is analogous to the AdS soliton for the $T^3$ case \cite{Horowitz:1998ha}.

 By doing a double analytic continuation on our static bubbles, we obtain  a black hole with topology $S^2\times S^1$. This should be viewed as a black string wrapping the $S^1$ of our solution with a noncontractible circle. It was shown in \cite{GSW}  that, under certain assumptions, such solutions could not exist but, as we will explain,  one of these assumptions is an asymptotic  condition which is not physically required and is not satisfied by our solutions.

An earlier attempt to describe gauge theory on  $S^2\times S^1$ involved a double analytic continuation of AdS and Schwarzschild AdS \cite{Birmingham:2002st,Ross:2004cb}. However, in that case the boundary metric was conformal to three dimensional de Sitter cross a circle, and   not the static space $S^2\times S^1\times R$.  Although this is not relevant to our main problem, we will review this in section 3 and point out a curious fact. The mass of the large bubble solution obtained from Schwarzschild AdS is 3/4 the expected Casimir energy of the weakly coupled field theory. A generalization of this solution yields a family of bubbles in which the minimum mass is precisely equal to the energy of the weakly coupled gauge theory. We do not understand the significance of this fact. 

Before we proceed to discuss the new solutions,  we note that five is the smallest bulk dimension in which the issue of new boundary topology arises.  In one lower dimension, the CFT lives in three dimensions and the spatial geometry can either be $S^2$ or $T^2$. Hence the dual description of the ground state is either global AdS or the the AdS soliton. However, in higher dimensions, there are even more possibilities. For $AdS_7$ examples, the dual CFT lives on a five dimensional space (cross time). Only for $S^5$ and $T^5$ is the dual of the ground state known. It would be interesting to consider other possibilities.

 \setcounter{equation}{0}

 \section{Static solutions}
 
 To describe the gravitational dual of gauge theory on $S^2\times S^1$, we
 consider static five dimensional spacetimes with a spatial $SO(3)\times SO(2)$ symmetry.  Without loss of generality, we can take the metric to be
\begin{equation} \label{metric}
ds^2 = - e^{\gamma(r)} dt^2 + \alpha(r) d\chi^2 + \frac{dr^2}{\alpha(r) \beta(r)} + r^2 d\Omega
\end{equation}
Einstein's equation in $AdS_{5}$ is
\begin{equation}
G_{a b} = - \Lambda g_{a b} = \frac{6}{l^2} g_{a b}
\end{equation}
 From the $G_{tt}$ component we obtain
\begin{equation} \label{Gtt}
r \beta' (4 \alpha + r \alpha') + 2 \beta(2 \alpha + 4 r \alpha' + r^2 \alpha'')  = 4 \Bigg( \frac{6 r^2}{l^2} + 1 \Bigg)
\end{equation}
while  the $G_{rr}$ component yields
\begin{equation} \label{Grr}
\gamma' = -\frac{1}{r} + \frac{\frac{4}{\beta} \Bigg( \frac{6r}{l^2} + \frac{1}{r}\Bigg) - 3 \alpha'}{4 \alpha + r \alpha'}
\end{equation}
The $G_{\chi \chi}$ component gives us
\begin{equation} \label{Gxx}
r(\alpha' \beta + \alpha \beta')(4 + r \gamma') + 4 \alpha \beta ( 1 + r \gamma') + r^2 \alpha \beta ({\gamma'}^2 + 2 \gamma'') = 4 \Bigg( \frac{6 r^2}{l^2} + 1 \Bigg)
\end{equation}
and from the  component on the $S^2$ we obtain
\begin{equation} \label{Gtheta}
r^2 \alpha' \beta' + r(2\alpha' \beta + \alpha \beta')(2 + r \gamma') + 2 r^2 \alpha'' \beta +2 r\alpha \beta \gamma' + r^2 \alpha \beta ({\gamma'}^2 + 2 \gamma'') =  \frac{24 r^2}{l^2}
\end{equation}
 Subtracting (\ref{Gtheta}) from (\ref{Gxx}) and substituting in (\ref{Grr}) and (\ref{Gtt}) to eliminate $\gamma'$ and $\beta'$ we find
\begin{equation} \label{beta}
\beta = \frac{ r \alpha' \Bigg( 1 + \frac{4 r^2}{l^2} \Bigg) - 8 \frac{r^2}{l^2} \alpha}{r^2 \alpha'^2 - \alpha (r\alpha' +r^2\alpha'')}
\end{equation}
Substituting (\ref{beta}) back into (\ref{Gtt}) we obtain a third order non-linear ODE for $\alpha$:
$$
64 r \alpha^3 \alpha' - 8(l^2 + 4 r^2) \alpha^2 \alpha'^2 + 4 r(l^2 - 10 r^2) \alpha \alpha'^3 + 4 r^2 (l^2 + 2 r^2) \alpha'^4$$
$$
 + \Big[ 8r (7r^2 - l^2) \alpha^2 \alpha' + r^2 (4r^2 - l^2) \alpha \alpha'^2 + 2 r^3(l^2 + 4 r^2) \alpha'^3 \Big] \alpha''
 $$
 $$
 - \Big[ 8 r^2 (l^2 + 3 r^2) \alpha^2 + 3 r^3 (l^2 + 4 r^2) \alpha \alpha' \Big] {\alpha''}^2
 $$
 \begin{equation} \label{ODE1}
 + \Big[-32 r^3 \alpha^3 + 4 r^2 (l^2 + 2 r^2)\alpha^2 \alpha' + r^3 (l^2 + 4 r^2) \alpha \alpha'^2 \Big] \alpha'''  = 0
 \end{equation}
  Once we have solved for $\alpha$,  we can obtain $\beta$ and $\gamma$ via (\ref{beta}) and (\ref{Grr}) respectively. The integration constant in $\gamma$ can be absorbed by rescaling $t$, so the solutions are uniquely determined by $\alpha$.  
  
There is a three parameter family of solutions to (\ref{ODE1}), but only a few are physically interesting. Suppose we want to specify initial data at the origin. Then  smoothness at the origin requires $\alpha'(0)=0$. Expanding around  $r=0$, (\ref{ODE1}) gives a constraint between $\alpha(0)$ and $\alpha''(0)$. This leaves a one parameter family of solutions. But (\ref{ODE1}) is invariant under constant rescaling of $\alpha$. This freedom is fixed by  requiring that the spacetime is asymptotically AdS, namely $\alpha = \frac{r^2}{l^2} + (\mathrm{lower \, \, \, order \, \, \, terms})$.  This produces a unique solution in which $\alpha$ is positive everywhere.  This solution has topology $R^4\times S^1$. Since the circle is topologically nontrivial, we can give $\chi$ any periodicity. So this solution exists for all $S^2\times S^1$ boundary conditions and is an AdS version of the standard Kaluza-Klein vacuum.

To construct a bubble solution in which the circle pinches off, we first pick a radius for the bubble, $r_0$, and require that $\alpha$ vanish there. Eq. (\ref{ODE1}) evaluated at $r=r_0$ now yields a constraint on $\alpha'(r_0)$ and $\alpha''(r_0)$. Fixing the overall scale again yields a unique solution for each $r_0$. So there is a unique static bubble for each radius.  For these bubble solutions the spacetime will be regular provided we periodically identify $\chi$ with period
\be \label{gencirclesize}
s = \frac{4\pi}{\alpha'(r_0) \sqrt{\beta(r_0)}}
\ee
We will see that $s$ is bounded from above and goes to zero for both very large and very small bubbles.  Thus these solutions exist only if the $S^1$ is small enough. If $r\ll l$, the cosmological constant terms in the field equations are negligible. Thus the bubbles with $r_0 \ll l$ look locally just like static bubbles in asymptotically flat spacetimes (i.e. time cross four dimensional Euclidean Schwarzschild) until $r\sim l$. Then the solution make a transition to their asymptotically AdS form.
    
 The asymptotic form of the solutions can be obtained analytically.  We take
  \be
  \alpha = \frac{r^2}{l^2} + f(r)
  \ee
  where for $r \gg l$, $f(r) \ll \frac{r^2}{l^2}$.  We also assume that  the first two derivatives of $f$ fall off like powers of $1/r$, i.e. $r^2 \vert f''(r) \vert \lesssim r \vert f'(r) \vert \lesssim \vert f(r) \vert$ for large $r$.   We can then solve the pair (\ref{beta}) and (\ref{Gtt}) to leading order, requiring also that $\beta\rightarrow 1$  as $r \rightarrow \infty$.\footnote{If one simply requires that $\beta$ remain finite at large $r$, the field equations ensure that $\beta\rightarrow 1$ asymptotically.} This determines $\alpha = \frac{r^2}{l^2} + \frac{1}{2} + \ldots$.  Then we define $\alpha = \frac{r^2}{l^2} + \frac{1}{2} + g(r)$ and continue perturbatively until we have as many terms as we desire.  We  find
  \begin{equation} \label{alph}
  \alpha = \frac{r^2}{l^2} + \frac{1}{2} + \Bigg[C_1 + \frac{1}{12}\log{\Big(\frac{r}{l}} \Big) \Bigg] \frac{l^2}{r^2} + \Bigg[C_2 - \frac{7}{216}  \log{\Big(\frac{r}{l}\Big)} \Bigg] \frac{l^4}{r^4} + \mathcal{O}\Bigg(\frac{l^6}{r^6}  \log\Big(\frac{r}{l}\Big) \Bigg)
\end{equation}
\be \label{bet}
\beta =  1 + \frac{l^2}{6 r^2} + \frac{1}{12}  \log{\Big(\frac{r}{l}\Big)}  \frac{l^4}{r^4}  - \Bigg[\frac{7}{216} + \frac{4 C_1}{3} + 6 C_2 \Bigg] \frac{l^4}{r^4} + \mathcal{O}\Bigg(\frac{l^6}{r^6}  \log\Big(\frac{r}{l}\Big) \Bigg)
\ee
\be
g_{rr} = \frac{l^2}{r^2} - \frac{2 l^4}{3 r^4} + \Bigg[\frac{85}{216} + \frac{C_1}{3} + 6 C_2 - \frac{1}{6}  \log\Big(\frac{r}{l}\Big) \Bigg]\frac{l^6}{r^6} + \mathcal{O}\Bigg(\frac{l^8}{r^8}  \log\Big(\frac{r}{l}\Big)  \Bigg)
\ee
\be 
g_{tt} = -\frac{r^2}{l^2} - \frac{1}{2} + \Bigg[\frac{4 C_1}{3} + 6 C_2 - \frac{11}{216} - \frac{1}{12} \log{\Big(\frac{r}{l} \Big)} \Bigg] \frac{l^2}{r^2}  + \mathcal{O}\Bigg(\frac{l^4}{r^4}  \log\Big(\frac{r}{l}\Big) \Bigg)
\ee
 where the higher order terms depend only on $C_1$ and $C_2$.  These two constants are not constrained by the asymptotic Einstein's equations and depend upon the interior.  A linear combination gives the mass of the solution while generic values of $C_1$ and $C_2$ will not correspond to a regular solution.  The logarithms in the expansion are somewhat surprising upon first exposure but they are allowed by the general Fefferman-Graham \cite{FeffGrahm} expansion.  
 
 It is interesting that for this $S^2\times S^1$ case,  the leading asymptotic behavior,    $\frac{r^2}{l^2} + \frac{1}{2}$, is half way between global AdS, $\frac{r^2}{l^2} + 1$, which is appropriate for the gauge theory on $S^3$, and AdS in Poincare coordinates, $\frac{r^2}{l^2} + 0$, which is appropriate for gauge theory on $R^3$.
 
 The metric functions for the solutions are qualitatively similar as we vary the  radius of the bubble,  so we just plot them for $r_0 = l$ in Figs. 1-3. For comparison, these plots also include the static solution with no bubble in which the circle does not pinch off. Note that  $g_{tt}$ is essentially the same for the two solutions.

  \begin{figure}[p]
\centering

\begin{picture} (0,0)
    	\put(-105,10){$g_{\chi \chi}$}
	      \put(125, - 135){$\frac{r}{l}$}
    \end{picture}

	\includegraphics[scale = 1]{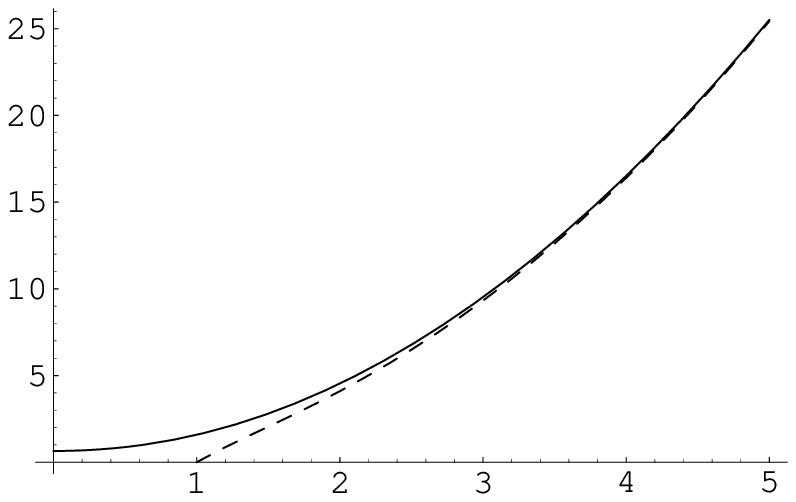}

\caption{$g_{\chi \chi}$ for $r_0 = l$ bubble (dashed line) and background (solid line)}
\end{figure}

\begin{figure}[p]
\centering
\begin{picture} (0,0)
    	\put(-110,5){$g_{rr}$}
         \put(120, - 137){$\frac{r}{l}$}
    \end{picture}

	\includegraphics[scale=1]{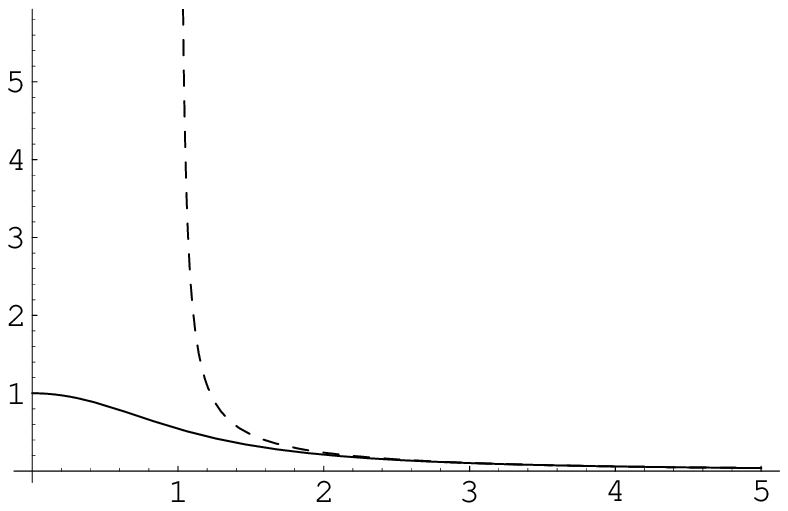}
	\caption{$g_{rr}$ for $r_0 = l$ bubble (dashed line) and background (solid line)}

 \end{figure}
 
  \begin{figure}[p]
\centering
\begin{picture} (0,0)
    	\put(-110,0){$g_{tt}$}
	      \put(120, -11){$\frac{r}{l}$}
    \end{picture}

	\includegraphics[scale = .7]{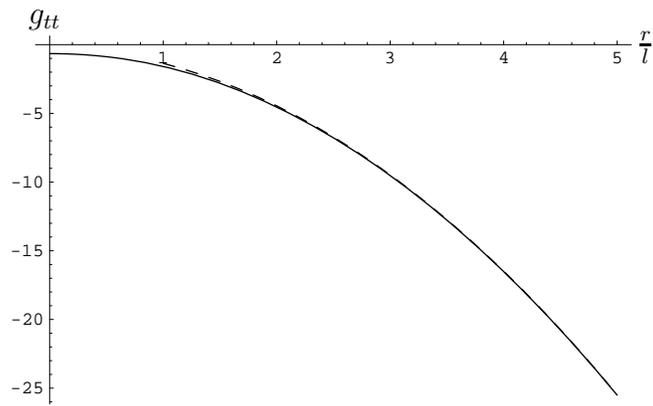}

\caption{$g_{tt}$ for $r_0 = l$ bubble (dashed line) and background (solid line) }
\end{figure}

 We now consider the energy of these solutions. There are various ways to define the mass of an asymptotically AdS spacetime (see, e.g., \cite{Abbott:1981ff,BK,AshtekarDas,Skendetal}). 
The existence of logarithms in the asymptotic metric make some of these  prescriptions divergent, although appropriate logarithmic counterterms can be included in a holographic definition of the energy.   Here we will use the background subtraction method using our solution with a noncontractible circle as the background.
 To be explicit, we adopt the definition of  \cite{HH} based on surface terms in the Hamiltonian
 \be
 E = -\frac{1}{8 \pi G} \int N (K - K_0)
 \ee
 where the integral is over a surface near infinity, $K$ the trace of the extrinsic curvature of that surface, and $K_0$ the trace of the extrinsic curvature of an isometric surface embedded in the  background spacetime.  Using the asymptotic expansions (\ref{alph}) and (\ref{bet}) one finds
 \be \label{Eexplic}
E = \frac{s l}{4 G} [ 5 (C_1 - K_1) + 18 (C_2 - K_2)]
 \ee
 where $K_1$ and $K_2$ are the values of the two constants $C_1, C_2$ in the reference background.
 These coefficients can be found by numerically matching the solution to (\ref{ODE1}) and its derivative to the asymptotic expansion (\ref{alph}) at large radius.  Using the solution with no bubble as a background one finds $K_1 \approx 0.107596$ and $K_2 \approx -0.033355$.\footnote{K. Skenderis has kindly informed us that using a counterterm subtraction prescription (choosing a renormalization scheme where the $h_4$ term is set to zero) one finds an energy for these metrics of the form (\ref{Eexplic})  with $-7/36$ replacing $-5 K_1 - 18 K_2$.  This gives a nonzero answer for the background which may be related to the  Casimir energy in the dual field theory.} 
 
  Table 1 lists the radii, circle size\footnote{The size of the circle in the bulk grows linearly with $r$. By ``circle size" we mean the periodicity of $\chi$, which is the circumference of the $S^1$ on the boundary in a conformal frame in which the $S^2$ has radius $l$.}, and mass of bubbles of various sizes.   For the sake of convenience in the table we take $l = 1$ and $G  = 1$.  Note that there is an upper bound on the circle size of  $s \approx 3.56 l$ in order for bubble solutions to exist.
For smaller size circles there are two bubble solutions (reminiscent of the Euclidean Schwarzschild solutions). Small bubbles have positive mass and large bubbles have negative mass. There is a zero mass bubble for 
$s \approx 2.98 l$.   These circle sizes tell us when we can first excite a transition to the bubble and when it can first occur spontaneously, respectively.

For large $r_0$, the static bubbles have $s\approx \pi l^2/r_0$ and $E\approx -\pi r_0^3/4Gl$. These are identical to the relations satisfied for the AdS soliton \cite{Horowitz:1998ha} (with the  rescaled volume $V_2$ of the two translationally invariant directions taken to be $4\pi l^2$). Thus, in a sense, our large static bubbles are approaching the AdS soliton. This is not literally true, since the asymptotic expansion of the metric is always different. A more precise statement is that the energy of the large bubble relative to the no-bubble solution approaches  the energy of the AdS soliton relative to AdS.

 \begin{table} \label{table1}
\vspace{.2in}
    \begin{center}
    \begin{tabular}{| l | c | r|}
    Bubble Size & Circle Size & Mass \\
    \hline
    $r_0 = 0.01$  & $ s \approx 0.1552178$ & ${E}  \approx 0.000252$ \\
    $r_0 = 0.1$  & $ s  \approx 1.445971$ &${E}  \approx 0.022194$ \\
    $r_0 = 0.2 $  & $ s  \approx  2.492528$ & ${E}  \approx 0.070755$ \\
    $r_0 = 0.3 $  & $ s  \approx 3.084494$ & ${E} \approx 0.118900$ \\
    $r_0 = 0.4 $  & $ s  \approx 3.325003$ & $E \approx 0.148915$ \\
      $r_0 = 0.4595 $  & $ s  \approx 3.562113$ & ${E}  \approx 0.15425612$ \\
      $r_0 = 0.4600 $  & $ s  \approx 3.5621509$& $E  \approx 0.15425688$ \\
      $r_0 = 0.4605 $  & $ s  \approx 3.5621508$& ${E}  \approx 0.15425687$ \\
     $r_0 = 0.5 $  & $ s  \approx 3.345232$ & ${E}  \approx 0.151825$ \\
     $r_0 = 0.6 $  & $ s  \approx 3.246060$ & ${E} \approx 0.123142$ \\
     $r_0 = 0.7 $  & $ s  \approx 3.090863$ & ${E}  \approx 0.059816$ \\
       $r_0 = 0.763762 $  & $ s \approx  2.979876 $ & ${E} \approx 4.256 \times 10^{-7}$ \\
       $r_0 = 0.7637625 $  & $ s  \approx 2.979875 $ & ${E} \approx -1.055 \times 10^{-7}$ \\
      $r_0 = 0.8 $  & $ s  \approx 2.915400$ & ${E}  \approx -0.041123$ \\
       $r_0 = 0.9 $  & $ s  \approx 2.738439$ & ${E} \approx -0.182980$ \\
      $r_0 = 1 $  & $ s  \approx 2.569128$ & ${E} \approx -0.369434$ \\
     $r_0 = 2 $  & $ s  \approx 1.487450$ & $E \approx -5.636213$ \\
      $r_0 = 3 $  & $ s  \approx 1.021741$ & ${E} \approx -20.32700$ \\
      $r_0 = 4 $  & $ s  \approx 0.774539$ & ${E}  \approx -49.14726$ \\
      $r_0 = 5 $  & $ s  \approx 0.622730$ & ${E}  \approx -96.81217$  \\
      $r_0 = 6 $  & $ s  \approx 0.520356$ & ${E} \approx -168.0359$ \\
       $r_0 = 7 $  & $ s  \approx 0.446753$ & ${E} \approx -267.5320$ \\
        $r_0 = 8 $  & $ s  \approx 0.391327$ & ${E}  \approx -400.0134$ \\
       $r_0 = 9 $  & $ s  \approx 0.348101$ & ${E}  \approx -570.1929$ \\
       $r_0 = 10 $  & $ s  \approx 0.313456$ & ${E}  \approx -782.7831$ \\
       $r_0 = 100 $  & $ s  \approx 0.031415$ & ${E} \approx -785373$ \\
        $r_0 = 10^3 $  & $ s  \approx 0.003142$ & ${E} \approx -7.85398 \times 10^{8}$ \\

    \hline
    \end{tabular}
    \caption{Bubble size, circle size, and mass of static bubbles ($l=1, G=1$)}
    \end{center}
    \end{table}
 
 For a large circle, the only static solution is our background which has zero energy (by definition). 
 For a small circle, the solution with the lowest mass is the large  bubble.   
One can now consider all initial data with conformal boundary $S^2\times S^1\times R$. If there is one which minimizes the mass, it  is expected to be static \cite{Sudarsky,Hertog:2005hm}.  Since AdS/CFT suggests that the energy is bounded from below, and we have found all static solutions with the expected symmetries, we are led to the following conjecture:

{\bf Conjecture}: Consider all five dimensional solutions to Einstein's equation with negative cosmological constant and conformal boundary $S^2\times S^1\times R$. Their energy  is always greater than the lowest energy of the static solutions  with the same size circle $s$.

To test this conjecture, one can consider  time symmetric initial data with $SO(3)\times SO(2)$ symmetry. In this case, the only equation that has to be satisfied is the scalar constraint (\ref{Gtt}). One can pick $\alpha$ arbitrarily and solve for $\beta$.  This gives a large class of bubble solutions.  We have examined a variety of such solutions, trying to find ones with arbitrarily negative energy analytically, or with energy smaller than the static solutions numerically. All such attempts have failed.  In particular, consider truncating $\alpha$ to its simplest expression consistent with the static asymptotics:
\be
\alpha(r) =  \frac{r^2}{l^2} + \frac{1}{2} +  \frac{1}{12}\log{\Big(\frac{r}{l}} \Big) \frac{l^2}{r^2} + C_1 \frac{l^2}{r^2}
\ee
Solving $C_1$ for $\alpha(r_0) = 0$ we find
\be\label{alphaconstr}
\alpha(r) =  \frac{r^2}{l^2} - \frac{r_0^4}{l^2 r^2} +  \frac{1}{2}\Big(1 -  \frac{r_0^2}{r^2}\Big) + \frac{1}{12}\log{\Big(\frac{r}{r_0}} \Big) \frac{l^2}{r^2} 
\ee
and hence $\alpha > 0$ for $r > r_0$.  As one varies $r_0$ one finds a continuous family of solutions. The free parameter in the solution for $\beta$ can be used to keep the size of the circle at infinity fixed. Let us consider in particular a circle size at infinity of $s = 1.021741l$,  which corresponds to a static solution with $r_0 = 3 l$.  The energies of the resulting solutions are shown in Fig. 4.  One can see that none of these solutions have energy less than the static bubble. The minimum energy  is approximately at $r_0 = 3.0015 l$ with a value $E \approx -20.32680 \frac{l^2}{G}$, approximately $10^{-3}$ percent higher than the static solution.
\begin{figure}[t]

\begin{picture} (0,0)
    	\put(-152,2){E}
         \put(177, -145){$\frac{r}{l}$}
    \end{picture}
    \centering

	\includegraphics[scale= 1]{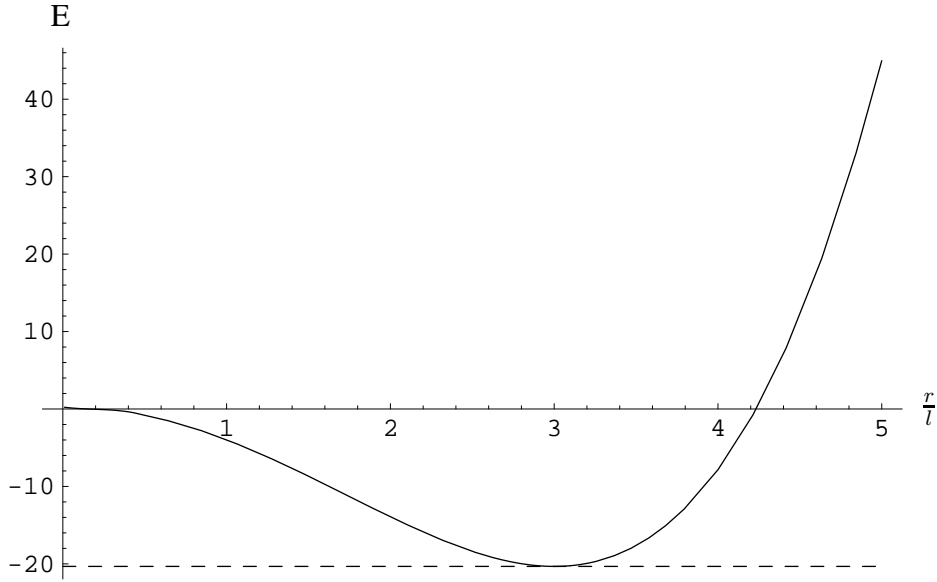}
	\caption{Energy for solutions (\ref{alphaconstr}) (solid line) and static solution (dashed line) for $s = 1.021741$}

\end{figure}

 If we take any static bubble of radius $r_0$ and analytically continue $\chi = i\tau$ and $t=iy$ we obtain a black hole solution
 \begin{equation} \label{bsmetric}
ds^2 =  - \alpha(r) d\tau^2 +  e^{\gamma(r)}dy^2 +\frac{dr^2}{\alpha(r) \beta(r)} + r^2 d\Omega
\end{equation}
The horizon is at $r=r_0$ and has topology $S^2\times S^1$. This is a black string wrapping the circle of our background. Since the inverse temperature of this black string is $s$, it is clear from Table 1 that the temperature is bounded from below, just like for spherical black holes in AdS. 
 
 It was shown in \cite{GSW} that there are no black holes in spacetimes with infinite fundamental group and negative cosmological constant, given what appears at first glance a minor restriction on the asymptotics.   The black strings we have described demonstrate that this restriction, while a useful technical condition for the arguments made in \cite{GSW},  is not required physically.  The previously assumed asymptotics (``condition S") stated the following. Let $h_{ij}$ be the metric on a static surface divided by $\alpha$. Consider the surfaces of constant $\alpha$ (for large $\alpha$). Then their extrinsic curvature (with respect to the metric   $h_{ij}$) cannot have eigenvalues of different sign. In our case this is an extra condition on the coefficients $C_1$ and $C_2$.  One can check that our black hole solutions, which are otherwise reasonable, simply do not satisfy it. In fact, there is an important physical reason why these black holes should exist: they describe the high temperature phase of the dual gauge theory on $S^2\times S^1$.
 
We conclude this section with a few  specifics about the numerical work. Recall that the equation for $\alpha$ (\ref{ODE1}) is scale invariant. For the solution in which  $\alpha$ never vanishes, 
 we can divide by $\alpha^4$ and reduce the third order ODE to a second order one by a substitution $\alpha = e^{\int dr  \lambda(r)}$.  In general, however, it is simpler to assume the form $\alpha =  \mu(r) e^{\int dr \nu(r)}$ where $\mu$ vanishes linearly at the bubble radius (provided a bubble is present) and goes as $\frac{r^2}{l^2}$ as $r \rightarrow \infty$ so that $\nu$ need not diverge.
 For bubbles not too large we define $u = \frac{r}{l}$ and take
 \be
 \alpha(u) = \mu(u) e^{\int_u^\infty dx \frac{\gamma(x)}{x^5}}
 \ee
 where
  \be
 \mu(u)=  u^2 + \frac{1}{2} - \Big( \frac{1}{2} + \frac{r_{0}^2}{l^2} \Big) \Big(\frac{r_0}{l u} \Big)^{10}
 \ee
 providing a smooth interpolation between the known behavior at large radius and small radius.  For large bubbles it is handy to  define $\epsilon$ such that
 \be
r = r_0 (1 + \epsilon)
 \ee
 so $\epsilon = 0$ corresponds to the bubble radius and for $\epsilon \gg 1$ we are at many times the bubble radius.  Then define
 \be
 \alpha(\epsilon) = \mu(\epsilon) e^{\int_\epsilon^\infty dx \frac{\omega(x)}{(1+ x)^5}}
 \ee
where
\be
\mu(\epsilon) =  \frac{r_0^2}{l^2} (1 + \epsilon)^2 + \frac{1}{2} - \Big( \frac{1}{2} + \frac{r_{0}^2}{l^2} \Big) \frac{1}{(1 + \epsilon)^{10}}
\ee
   For the no bubble solution we take
 \be
 \alpha(u) = \mu(u) e^{\int_u^\infty dx \frac{x}{1 + x^6} \nu(x)}
 \ee
 where
 \be
 \mu(u) = u^2 + \frac{1}{2} + \frac{ \log u}{12 u^2}\Big( 1 - \frac{7 }{18 u^2} \Big)\frac{u^{20}}{1 + u^{20}}
 \ee
The log term above incorporates the first two log terms in the asymptotic expansion of  $\alpha$.

 \setcounter{equation}{0}

 \section{Generalizing the Witten AdS bubble}
 
The AdS analog of Witten's expanding bubble solution can be obtained from a double analytic continuation of the Schwarzschild AdS metric and is given by  \cite{Birmingham:2002st,Ross:2004cb}
\be
ds^2 = \alpha(r) d\chi^2 + {dr^2\over \alpha(r)} + r^2 [ -d\tau^2 + \cosh^2\tau d\Omega]
\ee
where 
 \be\label{defalpha}
 \alpha = \frac{r^2}{l^2} + 1 - \frac{a_0^2}{r^2}
 \ee
 The bubble radius, $r_0$, is  the radius where $\alpha$ vanishes.
 The coordinate $\chi$ must be periodically identified  and we avoid conical singularities if we require the period to be
  \be \label{circlesize}
 s = \frac{2 \pi r_0 l^2}{2{r_0^2} +{l^2} }
 \ee
 
 The asymptotic metric is (conformal to) de Sitter cross a circle. Since this metric is time dependent, there is no conserved energy. We will consider the energy on the time symmetric surface $\tau=0$.
This energy, defined via background subtraction relative to the analytically continued AdS (i.e. $a_0 = 0$) is
  \be \label{Eschwarz}
 E = -\frac{s r_0^2 (l^2 + {r_0^2})}{4 l^3 G}
 \ee
Inverting (\ref{circlesize}) to find the size of the bubble for a given size circle at infinity, there are two solutions with radii
 \be
 r_0 = \frac{ l}{2s}\left[ \pi l  \pm \sqrt{\pi^2 l^2 -2{s^2} }\right]
 \ee
 provided $s \le {\pi l}/\sqrt{2}$, i.e. provided the circle size is small enough.  
 For small $s$, the small bubble has   $r_0 \approx s/2\pi $ and $E_{\mathrm{Schwarz}} \approx -{s^3}/{16 \pi^2 l G}$ whereas the large bubble has
\be
r_0 \approx \frac{\pi l^2}{s}
\ee
and
\be\label{adsbubble}
E_{\mathrm{Schwarz}} \approx - \frac{l^2 \pi^4}{4 G}  \Bigg(\frac{l}{s} \Bigg)^3
\ee
The bubble with  larger radius clearly has more negative energy.   We now consider a generalization of these bubbles.
 
 We wish to consider time symmetric initial data with a $S^1 \times S^2$ symmetry and hence may take the metric
 \be\label{initial}
 ds^2 =   \alpha(r) d\chi^2 + \frac{dr^2}{\alpha(r) \beta(r)} + r^2 d\Omega
 \ee
The only constraint (${}^4R =  -\frac{12}{l^2}$ and given explicitly in (\ref{Gtt})) is a first order equation for $\beta$ and hence for the same choice of $\alpha$ (\ref{defalpha}), there is a one parameter family of solutions for $\beta$
 \be
 \beta = 1 + \frac{c_1}{3 \frac{r^4}{l^4} + 2 \frac{r^2}{l^2} + \frac{1 - k^2}{3}} \Bigg( \frac{3 \frac{r^2}{l^2} + 1 - k}{3 \frac{r^2}{l^2} + 1 + k}\Bigg)^{\frac{1}{2 k}}
 \ee
 where $k = \sqrt{1 + \frac{3 r_0^2}{l^2} \Big( 1 + \frac{r_0^2}{l^2} \Big)}$.  We can then solve for this $c_1$ in terms of the size of the circle $\chi$ at infinity via (\ref{gencirclesize}) and hence obtain a solution for any size circle we like.
The energy of these solutions, again defined via background subtraction in comparison to continued AdS, is
\be \label{Eextend}
E = - \frac{r_0^2 s}{4 l G} \Bigg[ 1 + \frac{r_0^2}{l^2} + \frac{2}{3} \Bigg( \frac{3\frac{r_0^2}{l^2} + 1 + k}{3 \frac{r_0^2}{l^2} + 1 - k}\Bigg)^{\frac{1}{2 k}} \Bigg(\frac{4 \pi^2 r_0^2}{s^2 (2 \frac{r_0^2}{l^2} + 1)} - 1 - 2\frac{r_0^2}{l^2}\Bigg) \Bigg]
 \ee
In particular, let us consider the above solutions for small circles ($s \ll l$).  Let us examine them for small, intermediate, and large size ($r_0$) bubbles.  If $r_0 = \mathcal{O}(l)$ then $E = \mathcal{O}(\frac{l^3}{sG})$.  For small bubbles ($r_0 \ll l$)
\be
E \approx \frac{r_0 s}{3^{\frac{3}{2}}G} \Big(1 - \frac{4 \pi^2 r_0^2}{s^2} \Big)
\ee
which has a relative maximum at
\be
r_0 \approx \frac{s}{2 \pi \sqrt{3}}
\ee
with a value
\be
E_{\mathrm{max}} \approx \frac{s^2}{27 \pi G}
\ee
For large bubbles  ($r_0 \gg l$)
\be
E \approx \frac{r_0^2 s}{12 l G} \Big(  \frac{r_0^2}{l^2} - \frac{4 \pi^2 l^2}{s^2}\Big)
\ee
which has a minimum at
\be
r_0 \approx \frac{\sqrt{2} \pi l^2}{s}
\ee
and a minimum energy
\be
E_{\mathrm{min}} \approx -\frac{l^2 \pi^4}{3 G} \Bigg(\frac{l}{s} \Bigg)^3
\ee
This is clearly the minimum energy of all these solutions when $s$ is small. 
Intriguingly,  the energy of the analytically continued Schwarzschild solution (\ref{adsbubble})  is  ${3}/{4}$ the energy of the new solutions.  The energy of a large Schwarzschild bubble agrees with the energy of the AdS soliton which was shown in \cite{Horowitz:1998ha} to be precisely $3/4$ the Casimir energy of a weakly coupled ${\cal N}=4$ super Yang-Mills theory. This factor of $3/4$ has the same origin as the famous factor of $3/4$ relating the Hawking-Bekenstein entropy of a black three-brane to the entropy of the weakly coupled gauge theory \cite{Gubser:1996de}. The minimum energy of the new solutions is now seen to agree precisely with the Casimir energy of the gauge theory\footnote{The Casimir energy density of super Yang-Mills on $S^1 \times S^2$ should agree with $S^1 \times T^2$ when  $S^1$ is small.}. We do not know if this is just a coincidence or a hint of a deeper significance of these new solutions. Since there is a much larger family of solutions with time symmetric  initial data of the form (\ref{initial}) (obtained by varying $\alpha$ and solving the constraint for $\beta$) the one parameter family we have considered here do not appear  very special. In particular,  the solution with minimum energy in this family cannot be static because we saw in the last section that all static solutions have $g_{\chi\chi} \approx r^2 + 1/2 $  asymptotically.  However, this issue deserves further investigation.

Finally, we note that for fixed asymptotics, the new solutions have small positive mass for a small enough bubble (matching the expectation that for small bubbles the cosmological constant is negligible) while for very large bubbles the mass eventually becomes positive and large. This supports the idea that bubbles in AdS have a minimum mass (for fixed circle size). 

\vskip 1cm
\centerline{\bf Acknowledgments}
\vskip .5cm

It is a pleasure to thank J. Maldacena, D. Marolf,  R. Myers and J. Polchinski for discussions. This work was supported in part by NSF grant PHY-0244764.


\begin{thebibliography}{99}

\bibitem{Horowitz:1998ha}
  G.~T.~Horowitz and R.~C.~Myers,
  ``The AdS/CFT correspondence and a new positive energy conjecture for
  general relativity,''
  Phys.\ Rev.\ D {\bf 59} (1999) 026005
  [arXiv:hep-th/9808079].
  
  \bibitem{Witten:1998zw}
  E.~Witten,
  ``Anti-de Sitter space, thermal phase transition, and confinement in  gauge
  theories,''
  Adv.\ Theor.\ Math.\ Phys.\  {\bf 2} (1998) 505
  [arXiv:hep-th/9803131].
 
\bibitem{Surya:2001vj}
  S.~Surya, K.~Schleich and D.~M.~Witt,
  ``Phase transitions for flat adS black holes,''
  Phys.\ Rev.\ Lett.\  {\bf 86} (2001) 5231
  [arXiv:hep-th/0101134].
   
\bibitem{Aharony:2005ew}
  O.~Aharony, J.~Marsano, S.~Minwalla, K.~Papadodimas, M.~Van Raamsdonk and T.~Wiseman,
 ``The phase structure of low dimensional large N gauge theories on tori,''
  arXiv:hep-th/0508077.
   
\bibitem{Brill:1989di}
  D.~Brill and H.~Pfister,
  ``States Of Negative Total Energy In Kaluza-Klein Theory,''
  Phys.\ Lett.\ B {\bf 228}, 359 (1989).
  
  \bibitem{Brill:1991qe}
  D.~Brill and G.~T.~Horowitz,
  ``Negative energy in string theory,''
  Phys.\ Lett.\ B {\bf 262}, 437 (1991).
  
\bibitem{Witten:1981gj}
  E.~Witten,
  ``Instability Of The Kaluza-Klein Vacuum,''
  Nucl.\ Phys.\ B {\bf 195} (1982) 481.
  
  \bibitem{GSW}
  G.J. Galloway, S. Surya, and E. Woolgar. ``Non-Existence of Black Holes in Certain $\Lambda < 0$ Spacetimes,'' Class. Quant. Grav. 20 (2003) 1635-1648. [arXiv:gr-qc/0212079].
  
\bibitem{Birmingham:2002st}
  D.~Birmingham and M.~Rinaldi,
  ``Bubbles in anti-de Sitter space,''
  Phys.\ Lett.\ B {\bf 544} (2002) 316
  [arXiv:hep-th/0205246];
V.~Balasubramanian and S.~F.~Ross,
  ``The dual of nothing,''
  Phys.\ Rev.\ D {\bf 66} (2002) 086002
  [arXiv:hep-th/0205290];
  R.~G.~ Cai, ``Constant Curvature Black Hole and Dual Field Theory," Phys.\ Lett.\  {\bf B544} (2002) 176 [arXiv:hep-th/0206223].

 
\bibitem{Ross:2004cb}
  S.~F.~Ross and G.~Titchener,
  ``Time-dependent spacetimes in AdS/CFT: Bubble and black hole,''
  JHEP {\bf 0502} (2005) 021
  [arXiv:hep-th/0411128];
  V.~Balasubramanian, K.~Larjo and J.~Simon,
  ``Much ado about nothing,''
  Class.\ Quant.\ Grav.\  {\bf 22} (2005) 4149
  [arXiv:hep-th/0502111].
  
 \bibitem{FeffGrahm}
 C. Fefferman and C.R. Graham, Conformal invariants, in: \'Elie Cartan et les Mathematiques d'Aujourd'hui, Ast\'erisque, 1985, Numero hors series, Soc. Math. France, Paris, pp. 95-116.

  

\bibitem{Sudarsky}
D.~Sudarsky and R.~M.~Wald,
``Extrema of mass, stationarity, and staticity, and solutions to the 
Einstein
Yang-Mills equations,''
Phys.\ Rev.\ D {\bf 46}, 1453 (1992).

\bibitem{Hertog:2005hm}
  T.~Hertog and S.~Hollands,
  ``Stability in designer gravity,''
  arXiv:hep-th/0508181.

\bibitem{Abbott:1981ff}
  L.~F.~Abbott and S.~Deser,
  ``Stability Of Gravity With A Cosmological Constant,''
  Nucl.\ Phys.\ B {\bf 195} (1982) 76.
   
 \bibitem{BK}
 V. Balasubramanian and P. Kraus. ``A Stress Tensor for Anti-de Sitter Gravity,'' Comun. Math. Phys. 208 (1999) 413-428.  [arXiv:hep-th/9902121].
  
  \bibitem{AshtekarDas} 
  A. Ashtekar and S. Das, ``Asymptotically anti-de Sitter space-times: Conserved quantities,'' Class. Quant. Grav. {\bf 17},  L17 (2000) [arXiv:hep-th/9911230].

\bibitem{Skendetal}
M. Henningson and K. Skenderis, ``The Holographic Weyl anomaly" 
 JHEP {\bf 9807} (1998) 023
  [arXiv:hep-th/9806087];
S.~de Haro, K.~Skenderis and S.~N.~Solodukhin, ``Holographic Reconstrunction of Spacetime  and Renormalization in the AdS/CFT Correspondence," Commun.\ Math.\ Phys.\  {\bf 217} (2001) 595
[arXiv:hep-th/0002230];
   I. Papadimitriou and K. Skenderis, ``Thermodynamics of Asymptotically Locally AdS Spacetimes," JHEP {\bf 0508} (2005) 004
  [arXiv:hep-th/0505190];
 M.~C.~N.~Cheng and K.~Skenderis,
  ``Positivity of energy for asymptotically locally AdS spacetimes,''
  JHEP {\bf 0508} (2005) 107
  [arXiv:hep-th/0506123].
   
  
    
\bibitem{HH}
  S.~W.~Hawking and G.~T.~Horowitz,
  ``The Gravitational Hamiltonian, action, entropy and surface terms,''
  Class.\ Quant.\ Grav.\  {\bf 13} (1996) 1487
  [arXiv:gr-qc/9501014].
  
\bibitem{Gubser:1996de}
  S.~S.~Gubser, I.~R.~Klebanov and A.~W.~Peet,
  ``Entropy and Temperature of Black 3-Branes,''
  Phys.\ Rev.\ D {\bf 54} (1996) 3915
  [arXiv:hep-th/9602135].

\end{thebibliography}
\end{document}